\documentclass[prb,aps,showpacs,unsortedaddress,superscriptaddress,amsmath,amssymb,floatfix,twocolumn]{revtex4-1}
\usepackage{graphicx}% Include figure files
\usepackage{dcolumn}% Align table columns on decimal point
\usepackage{bm}% bold math
\usepackage[hidelinks,colorlinks=true,linkcolor=blue,citecolor=blue]{hyperref}
\usepackage[mathlines]{lineno}% Enable numbering of text and display math

\usepackage{braket}
\usepackage[usenames]{color}
\usepackage{float}
\usepackage{amsmath}

\begin{document}
\title{Role of strong correlation and spin-orbit coupling in $\textrm{LuB}_{4}$: a first principle study}
\author{Ismail Sk} 
\affiliation{Department of Physics, Bajkul Milani Mahavidyalaya, Purba Medinipur,West Bengal, 721655, India}
\affiliation{Department of Physics, Kazi Nazrul University, Asansol, West Bengal, 713340, India}

\author{Joydeep Chatterjee}
\affiliation{Department of Physics, Indian Institute of Technology Kharagpur, West Bengal, 721302, India.}
\author{A Taraphder}
\affiliation{Department of Physics, Indian Institute of Technology Kharagpur, West Bengal, 721302, India.}
\author{Nandan Pakhira $^2$}
%\affiliation{Department of Physics, Kazi Nazrul University, Asansol, West Bengal, 713340, India.}
%\date{\today}
%\maketitle
\begin{abstract}
The recent observation of magnetization plateaus in rare-earth metallic tetraborides has drawn a lot of attention to this class of materials. In this work, we investigate the electronic structure of one such 
canonical system $\textrm{LuB}_{4}$, using first-principle density functional theory, together with strong Coulomb correlation and spin-orbit coupling (SOC) effects. The electronic band structures show that 
$\textrm{LuB}_{4}$ is a non-magnetic correlated metal with completely filled $4f$ shell. The projected density of states (DOS) shows a continuum at the Fermi level (FL), arising mainly from hybridized Lu $d$ 
and B $p$ orbitals, along with some discrete peaks, well separated from the continuum. These peaks arise mainly due to core-level Lu $s$, $p$ and $4f$ atomic orbitals. Upon inclusion of SOC, the discrete peak 
arising due to Lu $p$ is split into two peaks with $j = 1/2$, $j = 3/2$ while the peak arising due to Lu $4f$ orbitals splits into two peaks with $j = 5/2$ and \,\, $j = 7/2$. These peaks will give rise to multiplet 
structure in core level X-ray photo-emission spectroscopy. Inclusion of strong correlation effects pushes the Lu $4f$ peak away from the FL while the qualitative features remain intact.
\end{abstract}
\pacs{}
\maketitle
\section{Introduction} 
 Compounds involving Boron like Boron Carbide (BC), $\textrm{MgB}_{2}$ ~\cite{article}, hexagonal Boron Nitride h-BN ~\cite{C7TC04300G}, $\textrm{SmB}_{6}$ and various metallic 
 tetra-borides display exotic mechanical and electronic attributes: very high tensile strength, high $\textrm{T}_{c}$ superconductivity, semi-metallic behavior with topological properties, Kondo insulating states, 
 magnetization plateaus etc. ~\cite{PhysRevB.91.155151,PhysRevB.88.180405,PhysRevX.3.011011}. These materials are under intense investigation due to their fascinating 
 properties like chemical stability, hardness, high melting points, low work function and magnetic ground states with geometric frustration. They have potential tehnological 
 applications such as narrow-band semiconductors ~\cite{CHEN2004L6,TAKEDA2004471,GAVILANO20001359}, thermoionic emission devices. Compounds like $\textrm{CrB}_{4}$ and $\textrm{MnB}_{4}$ 
 exhibit exceptional mechanical properties in the form of superior rigidity, ideal tensile and shear strength ~\cite{10.1063/1.3692777}.  

 The rare-earth metallic tetraborides with chemical formula $R\textrm{B}_{4}$ ($R$ being the rare-earth atom) have drawn interest due to recent observations of magnetization plateaus in $\textrm{TmB}_{4}$ 
 ~\cite{PhysRevB.93.174408}, $\textrm{ErB}_{4}$ ~\cite{Michimura} and $\textrm{NdB}_{4}$, $\textrm{HoB}_{4}$ ~\cite{Brunt}. In these systems the rare-earth atoms are arranged in a geometrically frustrated Shastry-Sutherland 
 lattice and show wide range of magnetic ground states ~\cite{SRIRAMSHASTRY19811069}. The $\textrm{CeB}_{4}$($\textrm{f}^{1}$) and $\textrm{YbB}_{4}$($\textrm{f}^{13}$) do not show any order, while $\textrm{PrB}_{4}$ is 
 ferro-magnetically ordered~\cite{article2}. On the other hand, $\textrm{TmB}_{4}$ and $\textrm{ErB}_{4}$ order anti-ferromagnetically at low temperatures ~\cite{ETOURNEAU1979531} in the presence of weak external magnetic 
 field. The role of spin-orbit coupling and strong Coulomb correlations on the electronic properties of $\textrm{RB}_{4}$ are thought to be key to understanding the novel and exotic properties observed in these systems. 

The strong correlation effects in $\textrm{RB}_{4}$ arise due to the presence of localized $4f$ orbitals in rare-earth ions. The trong correlation, present in various $3d$, \, $4d$ transition metal compounds as well as 
actinides with $5f$ orbitals, is central to the understanding of various novel and exotic properties observed in these systems. The valence state(s) of rare-earth atoms in $\textrm{RB}_{4}$ are found to be in di-, tri- and 
tetra-valent states in various rare-earth tetraborides ~\cite{goryachev1975electronic}. Recently, in $\textrm{YbB}_{4}$~\cite{PMID:35667370} fractional valance state of $\textrm{Yb}$ between $\textrm{Yb}^{2+}$ and 
$\textrm{Yb}^{+3}$ has also been reported.

In an earlier work~\cite{EPJB_IsmailSk,dae_ismail}, the role of strong spin-orbit coupling effects on the electronic properties of various rare-earth tetraborides in their non-magnetic ground states were 
investigated using first principles electronic structure methods as implemented in the Quantum Espresso (QE). One of the key features observed in various systems was the splitting of the Boron-derived peak and rare-earth 
$p$ orbitals into $j=1/2$ and $j=3/2$ multiplets. The splitting $\Delta E$ was found to be proportional to $Z^{n}$, where $Z$ is the atomic number of the rare-earth atoms and $n\approx 3.81$. However, the strong Coulomb 
correlation effects were neglected, and more importantly, the partial and total DOS had no contribution from $4f$ orbitals due to the absence of contributions from $4f$ orbitals towards the pseudo-potentials used in QE 
based calculations. 

In the present work we study the interplay of strong spin-orbit coupling and Coulomb correlation on the electronic structure of one such canonical system $\textrm{LuB}_{4}$, using first principles electronic structure 
methods as implemented in the Vienna ab initio simulation package (VASP)~\cite{KRESSE199615,PhysRevB.54.11169}. Most interestingly, the pseudo-potential used in VASP involves contribution from $4f$ orbitals of the rare-earth 
atom (Lu) as well. Also this system does not show any magnetic ordering due to closed shell atomic structure of Lu ($\textrm{4f}^{14}$), giving rise to vanishing of orbital and spin angular momentum. The organization of the 
rest of the paper is as follows: in Section II, we describe the crystal structure of $\textrm{LuB}_{4}$. In Section III we elaborate on the computational details for electronic band structure calculations. In Section IV, we 
discuss our results on the electronic band structure and density of states of $\textrm{LuB}_{4}$. Finally, in Section V we conclude.

% ==================================================================================================================================================

\section{Crystal structure}
\begin{figure}
\begin{center}
\includegraphics[scale=0.25]{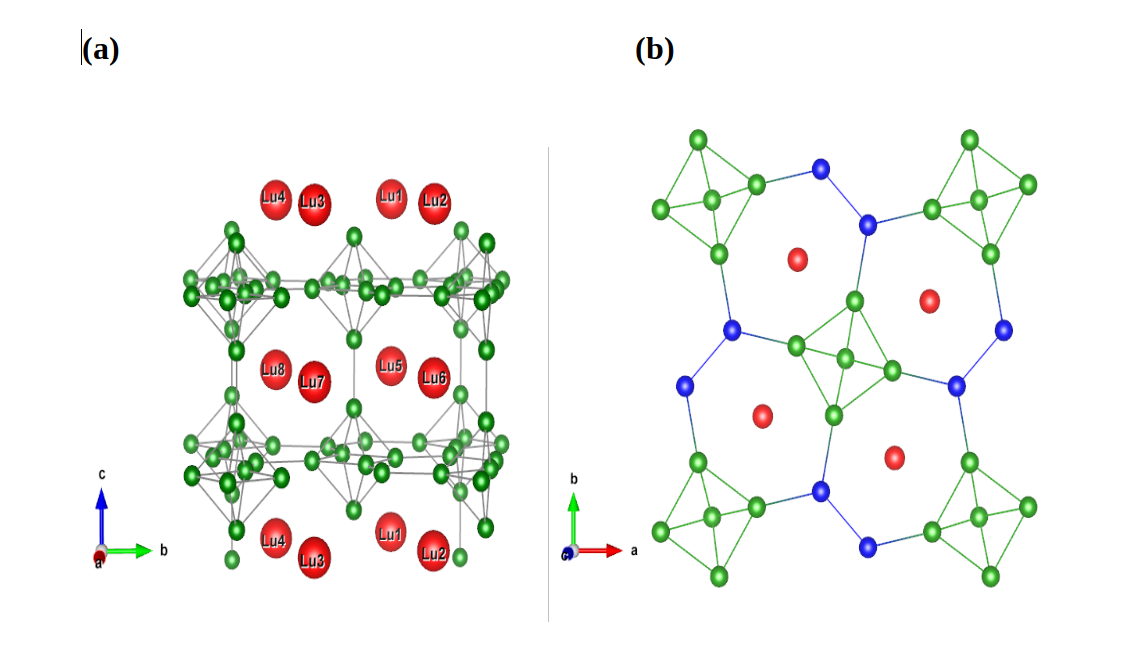} 	
\caption{(Color online) Tetragonal crystal structure of $\textrm{LuB}_{4}$. Panel (a) : Full structure consisting of different layers of rare-earth (Lu) ions (red) and B (green) stacked along $c$-axis. 
Panel (b) : Top view of the B (green and blue) sub lattice (in the $a-b$ plane) comprising of 7 atom ring and a square formed by the Lu (red) atoms.}
\label{Fig:CrystalStruct}
\end{center}
\end{figure}

$\textrm{LuB}_{4}$ has a tetragonal symmetry and belongs to space group P4/mbm~\cite{article3,FISK19811189}. In Fig.~\ref{Fig:CrystalStruct} we summarize the crystal structure of $\textrm{LuB}_{4}$ from different perspectives. 
Fig.~\ref{Fig:CrystalStruct}(a) displays the full tetragonal structure consisting of alternate layers of Lu and B ions stacked along $c$-axis. Fig.~\ref{Fig:CrystalStruct}(b)shows the top view of the crystal structure. There 
are two distinct types of B atoms - (i) planar and (ii) octahedral. Boron atoms form octahedra as well as 7-atom rings in the $a-b$ plane~\cite{https://doi.org/10.1002/adma.201604506}. Ring forming planar B atoms (shown in 
blue) which are not part of octahedra also forms dimers and these dimers are arranged in a regular pattern. It is important to mention that B atoms play a crucial role in the electronic structure as they are in the 
$sp$-hybridized state.
% ===============================================================================================================================================================================================================
% ===============================================================================================================================================================================================================

\section{Computational Details}
First-principles electronic structure calculations were performed using density functional theory (DFT)~\cite{PhysRev.136.B864,PhysRev.140.A1133} as implemented in VASP. All the calculations were performed with 
Perdew-Burke-Ernzerhof (PBE) exchange-correlation (XC) functionals in the framework of generalized gradient approximation (GGA)~\cite{PhysRevLett.77.3865} and linearized augmented plane wave (LAPW)~\cite{PhysRevB.59.1758} 
basis. The experimental lattice parameters and atomic positions of $\textrm{LuB}_{4}$ are taken from material project website~\cite{Jain2013}. The $\textrm{LuB}_{4}$ compound shows tetragonal lattice structure consisting of 
20 atoms in the unit cell. There are 4 Lu and 16 B atoms in each unit cell. For visualization and perodicity of the structure, VESTA ~\cite{Momma:db5098} software were used. The constituent atoms and their valence state are: 
$\textrm{Lu : [Xe]}4{f}^{14}5{d}^{1}6s^{2}$, $\textrm{B} : 2{s}^{2}2{p}^{1}$, respectively. The energy tolerance was set to $10^{-7}$ eV for structural relaxations. We have chosen a cut off energy $520$ eV throughout the 
calculations. All the structures were then fully relaxed by conjugate gradient (CG) algorithm when the difference in energy between succeeding iteration on each atom was less than $10^{-2}$ eV/atom. The standard Monkhorst-
Pack~\cite{PhysRevB.13.5188}(MP) $k$-grid was used for accurate integration over the irreducible Brillouin zone. $k$-mesh of size $11\times 11\times 11$ was used for relaxation and GGA self-consistent calculations whereas a 
denser $k$-meshe of size $13\times 13\times 13$ was used for the DOS calculations. The chosen $k$-meshes were good enough for well-converged calculations. In the GGA+U calculations, we have used $U=8.0$ eV and $J=1$ eV. 
These are typical values for $4f$ systems as reported in many calculations~\cite{PhysRevB.77.035135}. The electronic band structure was plotted along the path involving high symmetry points $\Gamma$=$(0,0,0)$, $X$ = ($\frac{\pi}{a},0,0)$, $M$=($\frac{\pi}{a}$,$\frac{\pi}{a}$,0), $Z$=(0,0,$\frac{\pi}{c}$), $R$=($\frac{\pi}{a}$,0,$\frac{\pi}{c}$), $A$=($\frac{\pi}{a}$,$\frac{\pi}{a}$, $\frac{\pi}{c}$). Calculated band structures were plotted along
the high symmetry directions $\Gamma-X-M-\Gamma$ and $Z-R-A-Z$.

%================================================================================================================================================================================================================= 
\section{Results and Discussions}
In Table\ref{tab1} we compare the lattice constants obtained from structural relaxation under GGA and GGA+U approximations to their observed experimental values. As can be clearly observed the calculated 
values are very close to the experimental values which validates the computed relaxed structures against the observed lattice structure. The calculated Fermi energy is tabulated in Table\ref{tab2}. It is clear 
that the Fermi energy is not much affected by the inclusion of either strong Coloumb correlation effect or spin-orbit coupling effects in this system.
% =================================================================================================================================================================================================================

\begin{table}[htbp]
\centering
\caption{Comparison of experimental and calculated (this work) lattice constants under various approximations}\label{tab1}%
\begin{tabular}{lcccccc}
\toprule
     \multicolumn{1}{c}{}&\multicolumn{2}{c}{Experimental} & \multicolumn{2}{c}{GGA} & \multicolumn{2}{c}{GGA+U} \\
 \hline  
    Materials & $a$ (\AA) & $c$ (\AA) & $a$ (\AA) & $c$ (\AA) & $a$ (\AA) & $c$ (\AA) \\
    \hline
    %\midrule
    $\textrm{LuB}_{4}$ & $7.02687$ & $3.96821$ & $7.03172$ & $3.97630$ & $6.99467$ & $3.94233$ \\
    \hline\hline
    %\bottomrule 
\end{tabular}
\end{table}

\begin{table}[htbp]
\centering
\caption{Calculated Fermi energy}\label{tab2}%
\begin{tabular}{lcccc}
\toprule 
& GGA  & GGA+SO & GGA+U & GGA+U+SO  \\
\hline
%\midrule
$\textrm{E}_{f}(eV)$ & $5.7625$ & $5.7567$ & $5.7694$ & $5.7638$ \\
\hline\hline

\end{tabular}
\end{table}

% =======================================================================================================================================================
%\pagebreak
\subsection{GGA and GGA+SO}
From the relaxed structure we calculate the self-consistent field (SCF) and subsequently the electronic band structure and density of states. Because of the completely filled $\textrm{4f}^{14}$ configuration 
of the Lu atom, $\textrm{LuB}_{4}$ is inherently non-magnetic. The band structure and DOS plotted in Fig. 2 and Fig. 4 also confirm the non-magnetic metallic state of $\textrm{LuB}_{4}$.
The Fermi level, indicated by red dashed line in both the plots, is set to zero. In Fig. 2(a) and 2(b) we show the band structure under GGA and GGA+SO approximations, respectively. It is clearly observed from 
Fig.2(a) that only four bands (indicated by numbers $ 1\cdots 4$) cross the Fermi level. This is similar to the reported band structure in $\textrm{TmB}_{4}$~\cite{pakhira2018electronic, PRADHAN20175532}. Also as reported 
earlier~\cite{PhysRevB.95.205140} for $\textrm{TmB}_{4}$, the bands near the symmetry point $M$ arises mainly due to hybridization between dimer B $2p$ orbital and Lu $4f$ orbital. In Fig. 2(b) we show the SOC effect on the 
electronic band structure. A noticeable feature is the splitting of the degenerate bands (labelled 5 and 6) at the $\Gamma$ point at an energy 1.5 eV. The large splitting gap is due to large atomic spin-orbit coupling present 
in Lu. Degeneracy-lifting is also present at various points along the $\Gamma - X - M - \Gamma$ directions. Along the $\Gamma-Z$ direction the detailed features of some of the bands (labelled 3 and 4) get modified in the 
presence of SOC. Along the $Z-R-A-Z$ direction and especially at the $A$ point, there is a large gap between the top and bottom bands. There are many flat bands arising due to deep core level states far away from the Fermi 
level, not shown in the figure. One additional band (orange, labelled 7) splits from band 3 and is found to cross the FL along the $\Gamma$-Z path in the presence of spin-orbit coupling. In Fig. 3(a) and (b) we have plotted 
the bands in a reduced energy scale and only along the shortened $\Gamma-Z$ path. The additional band split from band 3 in Fig. 3(a) is now clearly visible in Fig. 3(b).
% ==========================================================================================================================================================================================
\begin{figure}
\begin{center} 
\includegraphics[scale=0.35]{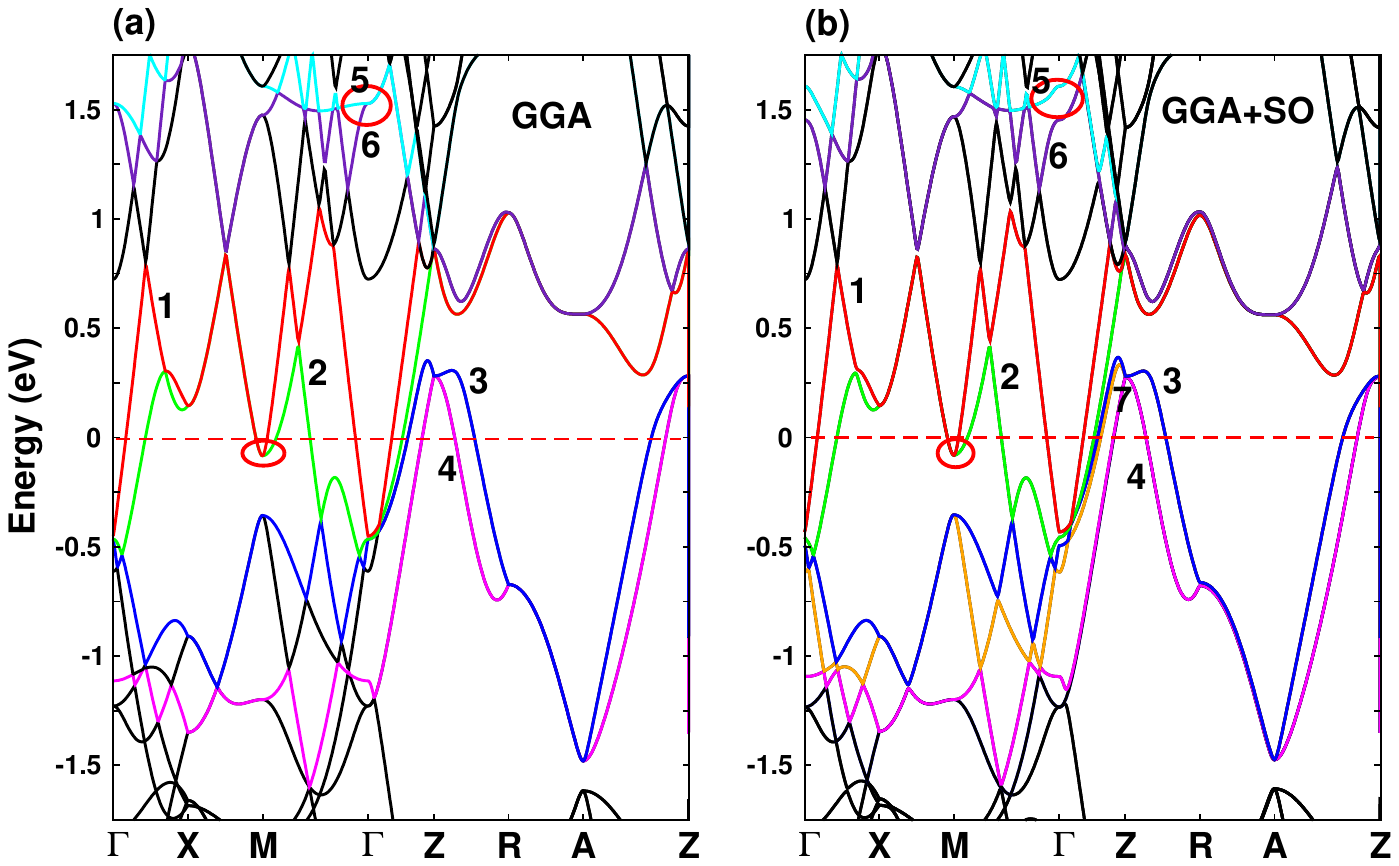}
\caption{(Color online)Panel : (a) Electronic band structure for $\textrm{LuB}_{4}$ under GGA approximation. Panel : (b) Electronic band structure under GGA+SO approximation. The Fermi level is set to zero.}
\label{Fig:BandStruct}
\end{center}
\end{figure}
% ===========================================================================================================================================================================================
\begin{figure}
\begin{center} 
\includegraphics[scale=0.3]{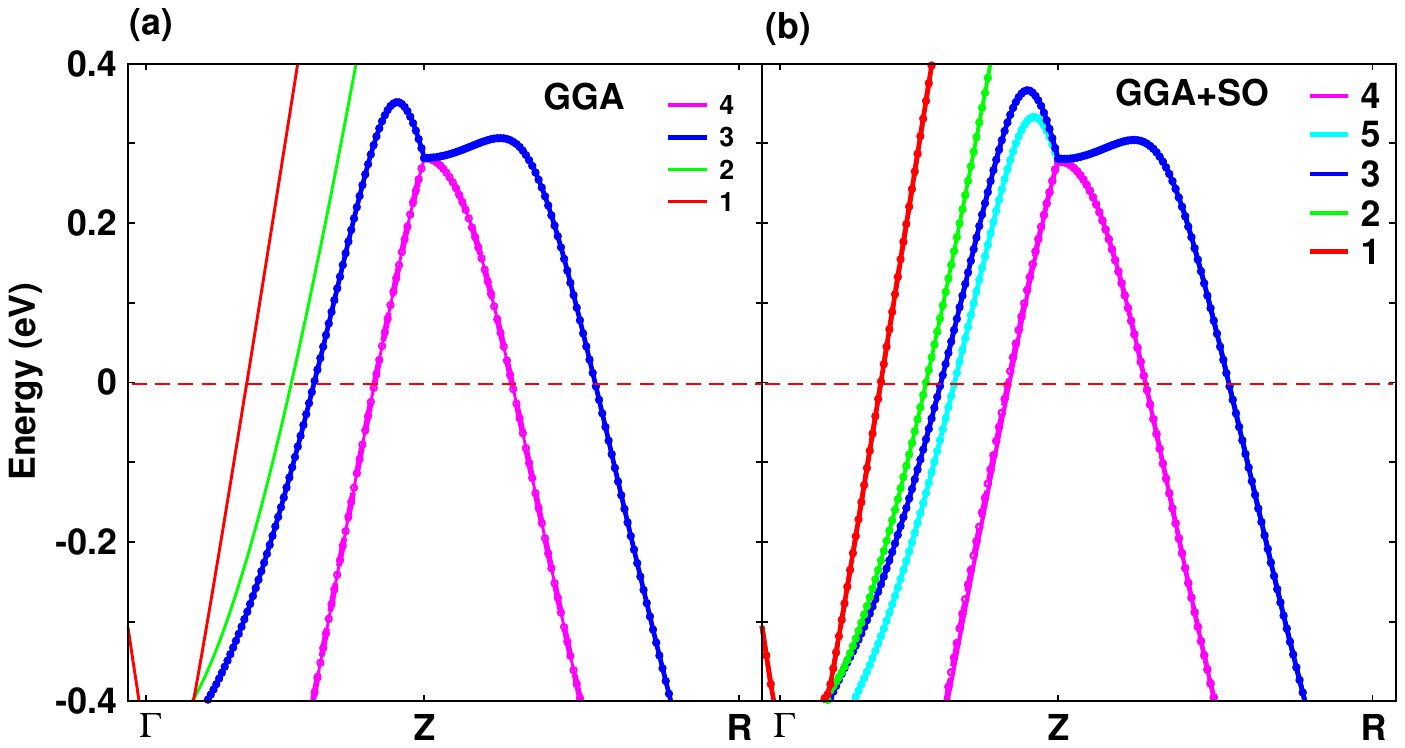}
\caption{(Color online)Panel: (a) and (b) Electronic band structure for $\textrm{LuB}_{4}$ under GGA and GGA+SO approximation plotted in reduced energy window and along the shortened path $\Gamma$-$Z$ path. 
	The band (labeled 3) in the left panel splits into two bands (labeled 3 and 5) in the right panel. }
\label{Fig:BandStruct}
\end{center}
\end{figure}
% ===========================================================================================================================================================================================
In Fig. 4 we plotted projected DOS (PDOS) as well as total DOS (TDOS) from various orbitals at a given site under GGA and GGA+SO approximations, respectively. In panel 4(a) and 4(b) are shown PDOS 
due to Lu and B, respectively, in the absence of SOC. As can be observed in panel 4(a), a discrete peak at -15.5 eV, well separated from the continuum DOS at Fermi level, arises due to B-$s$ and B-$p$ orbitals. 
Seen in panel 4(b), discrete large spectral peaks at -56.3 eV, -27.0 eV and -5.0 eV arise due to deep core level Lu-$s$, Lu-$p$ and Lu-$4f$ orbitals, respectively. The DOS at FL, as shown in the inset of panel 4(a), comes 
mainly from the hybridized B-$p$ and Lu-$d$ orbitals, with minor contributions from B-$s$, Lu-$s$ and Lu-$4f$ orbitals. 

In panel 4(c) and 4(d) we have shown the effect of SOC on PDOS due to Lu and B, respectively. As depicted in panel 4(c), except minor redistribution, the PDOS due to B largely remains unaffected. On the other hand, in panel 
4(b), the Lu-$p$ peak at -27.0 eV gets split into two peaks with $j=1/2$ and $j=3/2$, respectively. This feature was also observed in our earlier study~\cite{EPJB_IsmailSk,dae_ismail}. Additionally, the discrete peak due to 
$4f$ orbital of Lu atom gets split into two peaks with $j=5/2$ and $j=7/2$, respectively. The continuum density of states around the Fermi level gets enhanced in the presence of SOC, shown in the inset of panel 4(d). The 
enhancement is mainly due to Lu-$s$ and $d$ orbitals. However, the contribution from Lu-$4f$ orbital decreases. The top panels show the TDOS in each case. Due to the non-magnetic nature of $\textrm{LuB}_{4}$, TDOS for only one 
spin species is shown. 

% ========================================================================================================================================================
\begin{figure}
\begin{center}
\includegraphics[scale=0.35]{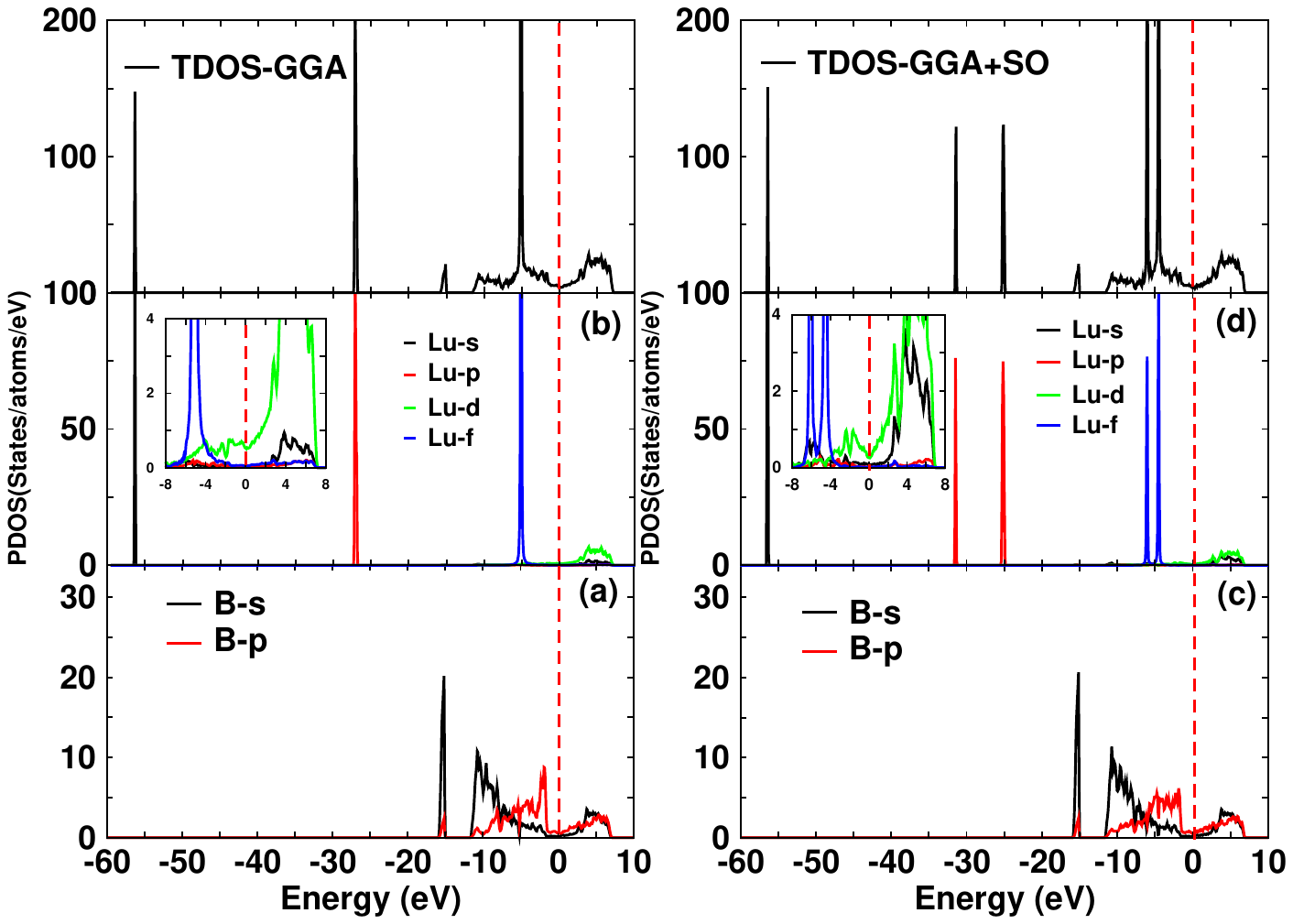} 
\caption{(Color online) Panel (a) and (b) : PDOS of $\textrm{LuB}_{4}$ in the absence of SOC. Contributions from different orbitals of B and Lu are shown. Discrete peaks at -56.3 eV, -27 eV and -5 eV arises 
mainly due to Lu s, Lu p and Lu f orbitals while the peak at -15.5 eV due to major B s with small contribution from B p orbital. Panel (c) and (d) : PDOS in the presence of spin-orbit coupling effect. The peak 
at -27 eV gets split into two peaks with $j = 1/2$ and $j = 3/2$ and another peak due to Lu-$4f$ orbital get split into two peaks with $j = 5/2$ and $j = 7/2$. Top left and right panel shows total dos from all 
orbitals under both of these approximations.}
\label{Fig:DOS}
\end{center}
\end{figure}
% ========================================================================================================================================================
\subsection{GGA+U and GGA+U+SO}
The correlation effect in rare-earth tetra-borides is quite significant and the electronic structure calculations should include these effects together with the strong SOC. In Fig. 5(a) and 5(b) we have summarized the 
electronic band structure of $\textrm{LuB}_{4}$ in the presence of $\textrm{U}$ and $\textrm{U+SO}$, respectively. As observed in Fig. 5(a), except an overall shift, the qualitative features of the energy bands near the 
Fermi level remain the same as in $U=0$. However, far away from the Fermi level, the qualitative features of some of the bands (like band 5 and 6) are different from the non-interacting case. It is important to mention 
that the effect of correlation, treated within the DFT so far, is a static mean-field effect. However, treated within the dynamical mean field theory (DMFT), it can drastically change the electronic band structure through 
the time-dependent self-energy functional. Inclusion of such effects is, for the time being, beyond the scope of this study. There is significant degeneracy-lifting at the $\Gamma$ point between band 5 and 6. Also, as in 
the earlier case, degeneracy lifting is visible at various points along the $\Gamma - X - M - \Gamma$ direction. However, there is degeneracy-lifting at an additional point ($Z$) at an energy $\sim 0.75$ eV. A detailed 
analysis also shows Lu-$4f-2p$ hybridization near $Z$. Inclusion of the SOC, as in $U=0$ case, gives rise to one additional Fermi level-crossing band (orange colour, labelled 7) shown in Fig. 5(b). 

As in the earlier case, in Fig. 6(a) and (b) we have plotted Fermi level- crossing bands along the shortened $\Gamma$-Z path in a reduced energy window under GGA+U and GGA+U+SO approximations, respectively. Due to the strong 
correlation effect, one of the bands (labelled 3, Fig. 6(a)) now nearly overlaps with another band (labelled 2). In the presence of strong SOC, as shown in Fig. 6(b), an additional Fermi level-crossing band (labelled 5) 
appears. However, at the Fermi level two bands (labelled 2 and 3) now overlap. This overlap is due to interplay between strong correlation and strong SOC effects.      

In Fig. 7 we have summarized the projected as well as total density of states for $\textrm{LuB}_{4}$ under GGA+U and GGA+U+SO approximations, respectively. In panel 7(a) we have shown PDOS due to B orbitals. 
As in the earlier case (GGA) a discrete peak at -15.5 eV, well separated from the continuum of density of states at the Fermi level, arises mainly due to B-$s$ deep core level states with minor contribution from 
B-$p$ orbital. The position of this peak is unaffected by the correlation effects. In panel 7(b) we have shown PDOS due to Lu orbitals. As in the earlier case three discrete peaks at -56 eV, -26.75 eV, and -8 eV 
arise mainly due to Lu-$s$, Lu-$p$, and Lu-$4f$ core levels states. The position of the peaks due to Lu-$s$ and Lu-$p$ slightly gets shifted. However, the position of the peak due to Lu-$4f$ orbital gets 
significantly modified. This is largely expected as the correlation effect is largest for localized $4f$ orbitals and the correlation effects on $p$, and $s$ orbitals are only through hybridization with $4f$ 
orbitals. In panel 7(c) and 7(d) we have shown the PDOS due to B and Lu orbitals under GGA+U+SO approximation, respectively. As in the earlier case, inclusion of SOC splits some of the discrete peaks while leaving 
other peaks largely unaffected. The peak at -26.75 eV arising due to the Lu-$p$ orbital gets split into two peaks with  $j = 1/2$ and $j = 3/2$, and another peak at -8 eV arising due to Lu-$4f$ orbital gets split into 
two peaks with  $j = 5/2$ and $j = 7/2$, respectively. Discrete peaks at -56 eV arising due to deep core level Lu-$s$ orbitals and at -15.5 eV arising due to B-$p$ and B-$s$ orbital remains unaffected.The DOS at 
the Fermi level gets enhanced due to Lu $4f$, $p$ and $d$ orbitals while the contribution from B orbitals largely remains the same.
% ==========================================================================================================================================================================================================
\begin{figure}
\begin{center}
\includegraphics[scale=0.35]{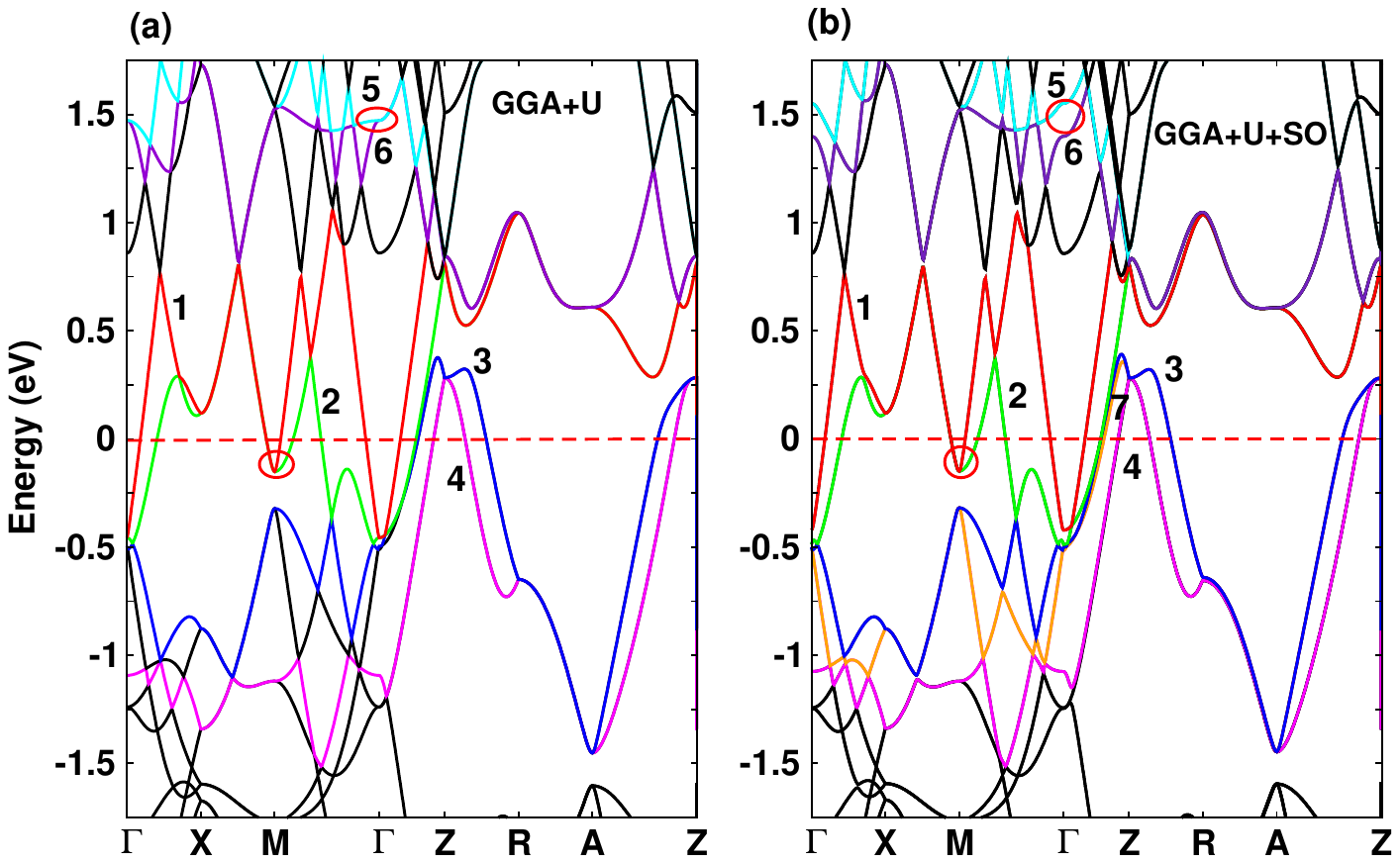} 
	\caption{(Color online)Panel : (a)  and (b) Electronic band structure for $\textrm{LuB}_{4}$ under GGA+U and GGA+U+SO approximations, respectively.}
\label{Fig:BandStruct}
\end{center}
\end{figure}
% ==========================================================================================================================================================================================================
% ==========================================================================================================================================================================================================
\begin{figure}
\begin{center}
\includegraphics[scale=0.35]{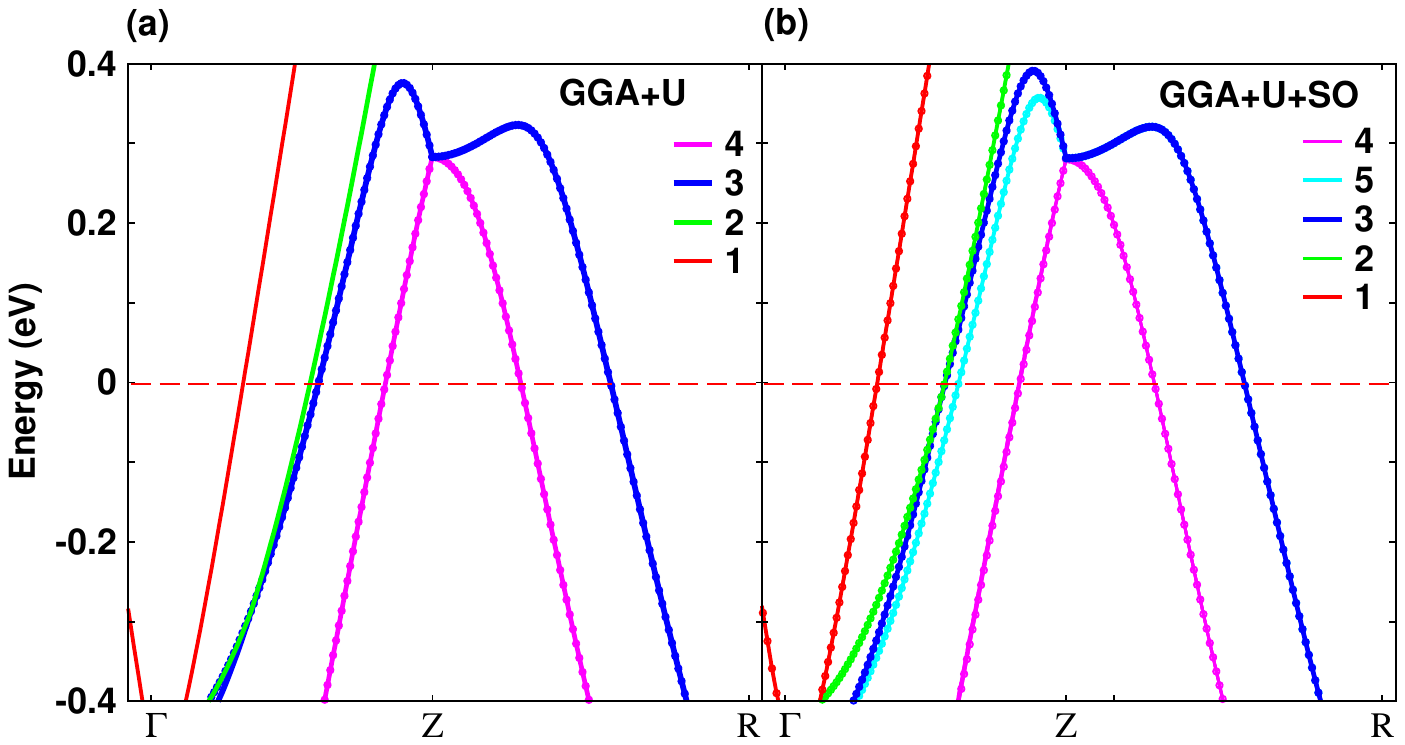} 
\caption{(Color online)Panel : (a) and (b) Electronic band structures for $\textrm{LuB}_{4}$ under GGA+U and GGA+U+SO approximations plotted in reduced energy window and along shortened $\Gamma$-Z path.}
\label{Fig:BandStruct}
\end{center}
\end{figure}
% ==========================================================================================================================================================================================================
% ==========================================================================================================================================================================================================
\begin{figure}
\begin{center}
\includegraphics[scale=0.35]{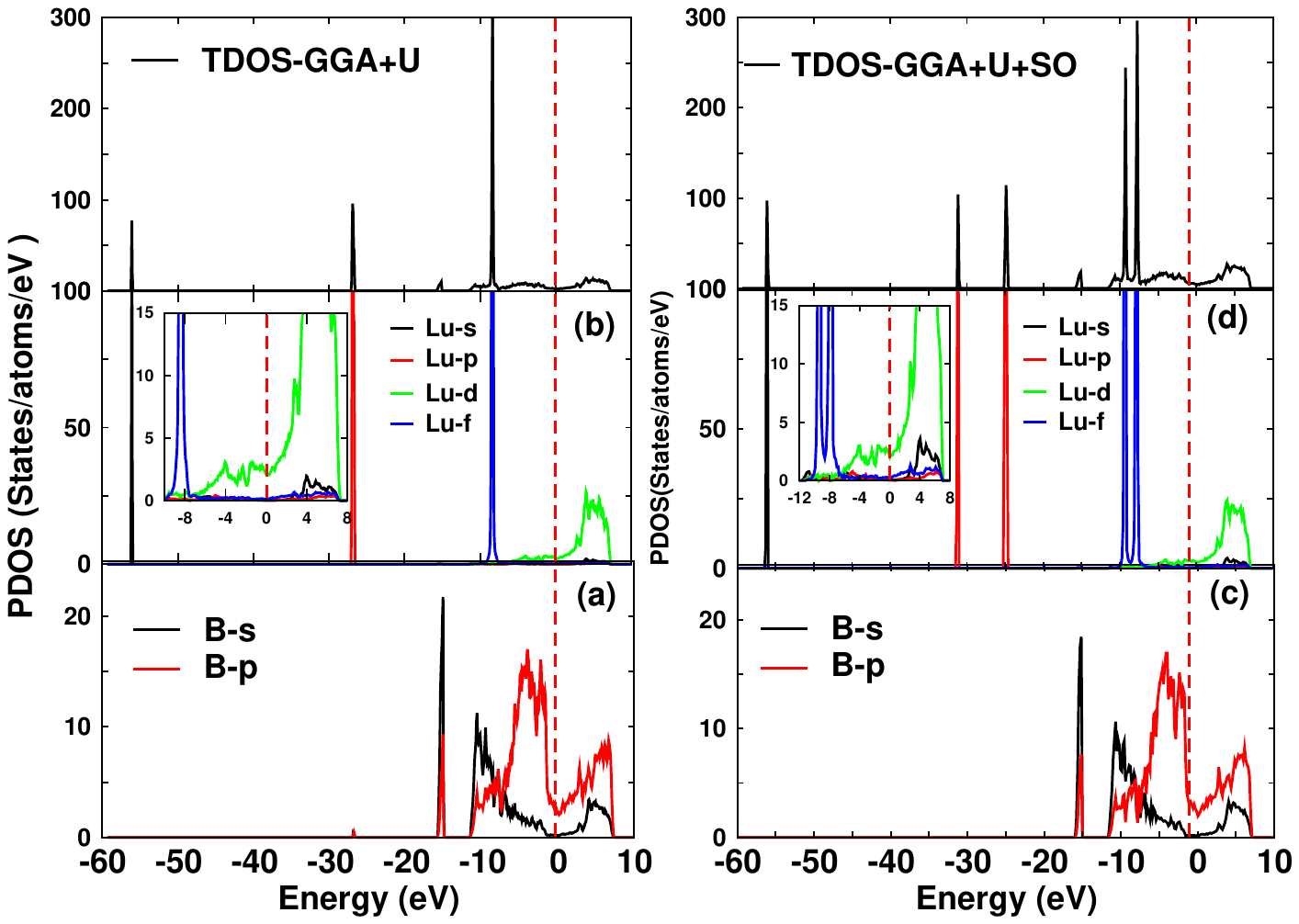} 
\caption{(Color online) Panel (a) and (b) : PDOS of $\textrm{LuB}_{4}$ under GGA+U approximation. Contributions from different orbitals of B and Lu are shown. Discrete peaks at -56 eV, -26.75 eV and -8 eV 
arises mainly due to Lu-$s$, Lu-$p$ and Lu-$4f$ orbitals while the peak at -15.5 eV arises due to major contributions from  B-$s$ with minor contribution from B-$p$ orbital. Panel (c) and (d): PDOS of 
$\textrm{LuB}_{4}$ with GGA+U+SO consideration. The peak at -26.75 eV gets split into two peaks with  $j = 1/2$ and $j = 5/2$ and another peak for Lu-$4f$ orbital get split into two peaks with  $j = 5/2$ and $j = 7/2$, 
respectively. The density of states at the Fermi level arises mainly due to hybridized Lu-$d$ and Lu-$4f$ orbitals with minor contributions from other orbitals. Top left and right panel shows TDOS from all 
the orbitals.}
\label{Fig:DOS}
\end{center}
\end{figure}
% ============================================================================================================================================================================================= 
\begin{figure}
\begin{center}
\includegraphics[scale=0.35]{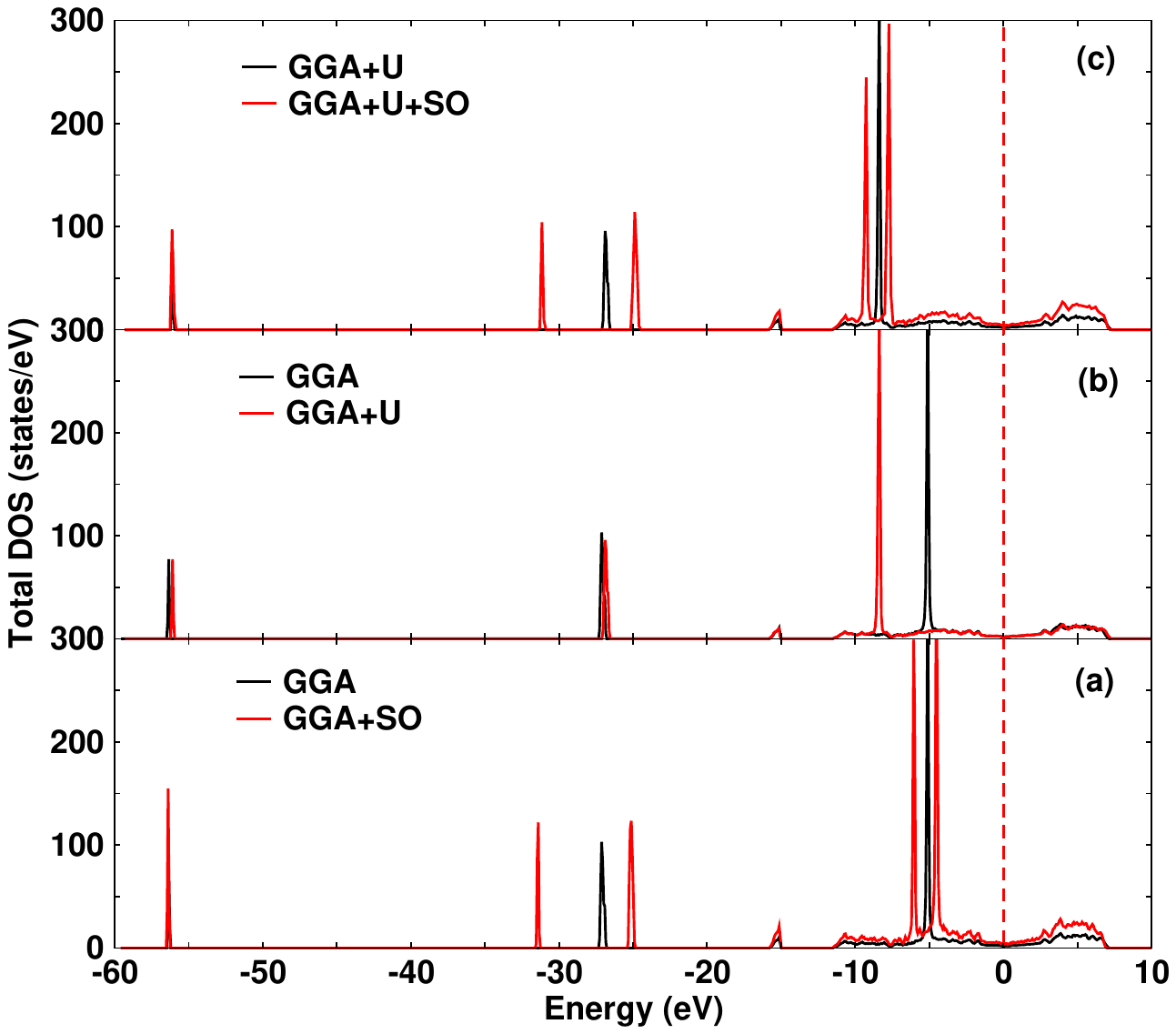} 
\caption{(Color online) TDOS of $\textrm{LuB}_{4}$ under GGA, GGA+U and GGA+U+SO approximations. Panel (a) : Comparison between GGA and GGA+SO approximations. The discrete peaks due to Lu-$p$ and Lu-$4f$ 
orbitals gets split into peaks characterized by  $j = 1/2$, $j = 3/2$ and  $j = 5/2$, $j = 7/2$, respectively. Panel (b) : Comparison between GGA and GGA+U approximations. The discrete peak due to Lu-$4f$ orbital gets 
shifted away from the Fermi level by nearly 3 eV while the shifts of other discrete peaks are less significant. Panel (c) : Comparison between GGA+U and GGA+U+SO approximations. Overall features are similar to 
case (a) but the position of various peaks are different due to strong correlation effects.}
\label{Fig:DOS}
\end{center}
\end{figure}
% ===============================================================================================================================================================================================
%\pagebreak
In Fig.8 we have summarized total density of states under various approximations. In panel 8(a) we have shown comparison between GGA and GGA+SO approximations. As mentioned earlier the discrete peak at -27 eV 
arising due to Lu-$p$ and $s$ orbitals gets split into two peaks with  $j = 1/2$ and $j = 3/2$ while the peak at -5 eV arising due to Lu-$4f$ orbital gets split into two peaks with  $j = 5/2$ and $j = 7/2$. The density 
of states at and around the Fermi level enhances. However total integrated spectral weight under GGA from both the spin channels and under GGA+SO approximations remain the same. In panel 7(b) we have shown comparison 
between GGA and GGA+U. The density of states at the Fermi level does not change but the discrete peak at -5 eV shifts to -8 eV due to correlation effects. The shape of the peak remains same. Further, as shown in 
panel 8(c), upon turning on the SOC effect, the discrete peaks at -8 eV and at -27 eV gets split into peaks characterised by  $j = 5/2$,  $j = 7/2$ and  $j = 1/2$, $j = 3/2$ respectively.

The splitting of various peaks arising due to strong SOC effect will be directly visible as detailed multiplet structure in core-level X-ray photo-emission spectroscopy. However, various non-local screening effects not 
included under GGA approximations can broaden various peaks. These non-local screening effects can be included under various non-local approximations beyond GGA like GW scheme and beyond the scope of the present study.  

\section{Conclusions}
To summarise, we have studied the electronic structure of $\textrm{LuB}_{4}$ under strong correlation and spin-orbit coupling. The electronic band structure under GGA shows 4 Fermi level-crossing bands as reported earlier for $\textrm{TmB}_{4}$. Inclusion of strong SOC under GGA+SO approximation shows strong degeneracy-lifting at $\Gamma$ point between two bands (labelled 5 and 6). Also, the degeneracy-lifting 
is visible at various other points along the $\Gamma - X - M - \Gamma$ direction. Most interestingly, in the presence of SOC, an additional band appears to cross the Fermi level along the $\Gamma$-$Z$ direction. 
The projected density of states shows discrete peaks at -56 eV, -27 eV and -5 eV arising due to Lu-$s$, Lu-$p$ and Lu-$4f$ orbitals, respectively. The density of states at the Fermi level is due to hybridized Lu-$p$, 
$s$ and $4f$ orbitals. Inclusion of strong correlation effects through local $U$ causes the shifting of -5 eV peak further away from the Fermi level while the other features remains nearly same. Upon inclusion of 
strong SOC effects the peaks at -8 eV gets split into two peaks characterized by  $j = 5/2$, $j = 7/2$ while the peak at -27 eV gets split into two peaks characterized by $j=1/2$, $j=3/2$. Also, the density of states at and 
around the Fermi level shows enhancement.

The present study will be extended to other rare-earth tetra-boride systems with magnetically ordered ground states to understand the interplay of strong correlation, strong SOC with magnetism in these fascinating systems. Also, the dynamic correlation effects, not included under DFT approximations, are extremely important and will be included in a future study using dynamical mean field theory (DMFT) as implemented under LDA+DMFT approximation.

\section*{Acknowledgements}
 This work is partially supported by WB-DSTBT research grant no. STBT- 11012(26)/31/2019-ST SEC. One of us (NP) would like to acknowledge hospitality of IIT, Kharagpore where part of the manuscript was written. 
\bibliographystyle{unsrt}
\bibliography{Ref}
%\markboth{References}{}
   %\addcontentsline{toc}{chapter}{References}
%\label{page:References}
\end{document}